\documentclass[a4paper,fleqn,usenatbib]{mnras}

\usepackage{newtxtext,newtxmath}

\usepackage[T1]{fontenc}

\usepackage{graphicx}	
\usepackage{amsmath}	
\usepackage{amssymb}	

\title[Resolving the Centrifugal Barrier in L1527]{Vertical Structure of the Transition Zone from Infalling Rotating Envelope to Disk in the Class 0 Protostar, IRAS04368+2557}

\author[N. Sakai et al.]{
Nami Sakai,$^{1}$\thanks{E-mail: nami.sakai@riken.jp}
Yoko Oya,$^{2}$
Aya E. Higuchi$^{1}$
Yuri Aikawa,$^{3}$
Tomoyuki Hanawa,$^{4}$
\newauthor
Cecilia Ceccarelli,$^{5}$
Bertrand Lefloch,$^{5}$
Ana L\'{o}pez-Sepulcre,$^{2,6}$
Yoshimasa Watanabe,$^{2}$
\newauthor
Takeshi Sakai,$^{7}$
Tomoya Hirota,$^{8}$
Emmanuel Caux,$^{9,10}$
Charlotte Vastel,$^{9,10}$
Claudine Kahane,$^{5}$
\newauthor
and Satoshi Yamamoto$^{2}$
\\
$^{1}$The Institute of Physical and Chemical Research (RIKEN), 2-1, Hirosawa, Wako-shi, Saitama 351-0198, Japan\\
$^{2}$Department of Physics, The University of Tokyo, Bunkyo-ku, Tokyo 113-0033, Japan\\
$^{3}$Center for Computational Science, University of Tsukuba, Tsukuba, Ibaraki 305-8577, Japan\\
$^{4}$Center for Frontier Science, Chiba University, Chiba, 263-8522, Japan\\
$^{5}$Universite Joseph-Fourier, Laboratoire d'Astrophysique de Grenoble BP 53X 5 rue des Moutonn\'{e}es Grenoble Cedex 9, FR 38042\\
$^{6}$Institut de Radioastronomie Millim\'{e}trique 300 rue de la Piscine, Domaine Universitaire 38406 Saint Martin d'H\'{e}res, France\\
$^{7}$Department of Communication Engineering and Informatics, Graduate School of Informatics and Engineering, The University of Electro-Communications, \\
Chofugaoka, Chofu, Tokyo, 182-8585, Japan\\
$^{8}$National Astronomical Observatory of Japan, Osawa, Mitaka, Tokyo 181-8588, Japan\\
$^{9}$Universite de Toulouse, UPS-OMP, IRAP, Toulouse, France
$^{10}$CNRS, IRAP, 9 Av. Colonel Roche, BP 44346, F-31028 Toulouse Cedex 4, France
}

\date{Accepted XXX. Received YYY; in original form ZZZ}

\pubyear{2016}

\begin{document}
\label{firstpage}
\pagerange{\pageref{firstpage}--\pageref{lastpage}}
\maketitle

\begin{abstract} 
We have resolved for the first time the radial and vertical structure of the almost edge-on envelope/disk system of the low-mass Class 0 protostar L1527. For that, we have used ALMA observations with a spatial resolution of 0.25$^{\prime\prime}$$\times$0.13$^{\prime\prime}$ and 0.37$^{\prime\prime}$$\times$0.23$^{\prime\prime}$ at 0.8 mm and 1.2 mm, respectively. The L1527 dust continuum emission has a deconvolved size of 78 au $\times$ 21 au, and shows a flared disk-like structure.  A thin infalling-rotating envelope is seen in the CCH emission outward of about 150 au, and its thickness is increased by a factor of 2 inward of it. This radius lies between the centrifugal radius (200 au) and the centrifugal barrier of the infalling-rotating envelope (100 au). The gas stagnates in front of the centrifugal barrier and moves toward vertical directions. SO emission is concentrated around and inside the centrifugal barrier.  The rotation speed of the SO emitting gas is found to be decelerated around the centrifugal barrier.  A part of the angular momentum could be extracted by the gas which moves away from the mid-plane around the centrifugal barrier.  If this is the case, the centrifugal barrier would be related to the launching mechanism of low velocity outflows, such as disk winds.  
\end{abstract}




\section{Introduction}
\begin{table}
	\centering
	\caption{List of Observed Molecules$^{\mathrm{a}}$}
	\label{tab:line}
	\begin{tabular}{lccc} 
		\hline
		\multicolumn{1}{c}{Transition}&Frequency [GHz]&S$\mu^2$&$E_u^{\mathrm{b}}$ [K] \\
		\hline
		\multicolumn{4}{c}{CCH} \\
		\hline
		$N=4-3$, $J=9/2-7/2$,&$\quad$&$\quad$&$\quad$ \\
		$F=5-4$&349.3377056&2.8987&41.91 \\
		$F=4-3$&349.3389882&2.3163&41.91 \\
		$N=3-2$, $J=7/2-5/2$,&$\quad$&$\quad$&$\quad$ \\
		$F=4-3$&262.0042600&1.711&25.15 \\
		$F=3-2$&262.0064820&1.633&25.15 \\
		\hline
		\multicolumn{4}{c}{SO} \\
		\hline
		$J_N=7_8-6_7$&340.7141550&16.244&81.25 \\
		$J_N=7_6-6_5$&261.8437210&16.383&47.55 \\
		\hline
	\end{tabular}
\begin{list}{}{}
\item[$^{\mathrm{a}}$] Taken from CDMS (M$\ddot{\rm u}$ller et al. 2005).
\item[$^{\mathrm{b}}$] Upper state energy.
\end{list}
\end{table}
In previous ALMA Cycle 0 observations, we imaged, on a 100~au scale, the distribution of several molecules around the Class 0/I protostar, IRAS04368+2557, in L1527. The images revealed an infalling-rotating envelope (IRE) system with an edge-on geometry \citep[e.g.][]{sak14a, oya15}. This IRE is traced by CCH, c-C$_3$H$_2$, and CS lines. Its centrifugal barrier ($r_{\rm CB}$; a half of the centrifugal radius, $r_{\rm CR}$) was clearly identified at a radius of ~100 au by analyzing the IRE kinematics with those lines \citep{sak14a,sak14b}. SO emission shows a ring-like structure with the radius of the centrifugal barrier. The enhanced SO emission is probably due to the liberation of SO and other sulfur-bearing molecules from dust grains by a weak accretion shock.  On the other hand, H$_2$CO likely traces IRE, a shocked zone around $r_{\rm CB}$, and the inner disk.  Thus, physical changes in the vicinity of the protostar are exquisitely highlighted by chemical changes.  A simple ballistic model of the IRE beautifully reproduces the observed kinematic structure \citep{sak14a,sak14b,oya15}.  The inclination angle of the IRE is determined to be 85$^\circ$ (90$^\circ$ for the edge-on), and the total mass of the protostar including the inner disk is 0.18 M$_\odot$.  The ALMA observations of C$^{18}$O and SO were also reported by \cite{oha14}.

The centrifugal barrier was then identified in a few other young stellar objects. Note that different chemical tracers are used depending on the source chemical composition: carbon-chain molecules in the case of WCCC (warm-carbon-chain-chemistry) protostars (L1527, IRAS15398-3359, and TMC-1A) and complex-organic-molecules in hot corinos (IRAS16293-2422) \citep{sak13,sak14a,sak14b,sak16,oya14,oya16}. An possible indication of the presence of the centrifugal barrier, such as a SO ring structure, was also reported in other low-mass protostellar sources \citep[e.g.][]{yen14,pod15,lee16}.  It is generally thought that the disk radius is close to the centrifugal radius, where the gravitational force is balanced by the centrifugal force.  However, high-angular-resolution observations have revealed that this expectation is too simplistic.  The envelope gas, at least a fraction of it, keeps infalling beyond the centrifugal radius into the centrifugal barrier, at least in the above sources.

This transition zone from an IRE to a rotationally-supported disk (Keplerian disk) was unexpected, and its finding is one of the breakthroughs obtained by high resolution molecular line observations of ALMA.  Thus, the detailed study of the transition zone around the centrifugal barrier is mandatory in understanding disk formation. Here we report detailed distributions of the gas around the centrifugal barrier by resolving radial and vertical structures of the disk/envelope system in L1527.

\section{Observations}
The observations were carried out on 2015 July 18 (Band 6) and July 18 and 20 (Band 7) in the C34-7 ALMA Early Science configuration.  They used 34 twelve-meter antennas for Band 6, and 37-39 antennas for Band 7.  The perceptible water vapor during the observations was 0.45 mm.  The data were obtained over a baseline range of 19.5-1158.0 m for Band 6, and 33.0-1565.0 m for Band 7, so that the emission extended more than 5.6$^{\prime\prime}$ and 2.4$^{\prime\prime}$ for Band 6 and Band 7, respectively, is solved out.  The field centre is $(\alpha_{2000}, \delta_{2000}) = (04^{\rm h} 39^{\rm m} 53^{\rm s}.87, 26^{\circ} 03^{\prime} 09^{\prime\prime}.6)$, which is the 0.8~mm continuum peak of L1527 observed with ALMA in Cycle 0 \citep{sak14a}.  The primary beam (half power beam width) is 22.92$^{\prime\prime}$ and 17.63$^{\prime\prime}$ at Band 6 and Band 7, respectively.  The total integration time and the typical system temperature were 40.95 min, 58.06 min and 60-100, 100-200 K for Band 6 and Band 7, respectively.  The backend correlator was split into 16 small windows, each of which has the bandwidth and channel spacing of 59 MHz and 61 kHz, respectively.  The observed lines are summarized in Table~\ref{tab:line}.  Phase calibration was conducted every 10 minutes by using J0510$+$1800 for Band 6 and J0438$+$3004 for Band 7.  J0423-0120 was used for bandpass calibration, while J0433$+$0521 was used for a part of the Band 7 data.  The absolute flux density scale was derived from J0423-013 and J0510$+$1800.  The data calibration was performed in the antenna-based manner, and uncertainties are less than 10 \%.  The continuum image was prepared by averaging line-free channels.  Line images were obtained after subtracting the continuum directly from the visibilities.  After the normal calibration, self-calibration was applied by using the continuum data, where the continuum peak was set to be the field-centre position.  The synthesized beam size for the continuum map in Band 7 is $0.26^{\prime\prime} \times 0.13^{\prime\prime}$ with the position angle (PA) of -27.0$^\circ$.

\section{Results and Discussions}

\subsection{Continuum}

\begin{figure*}
   \centering
   \includegraphics[height=10cm]{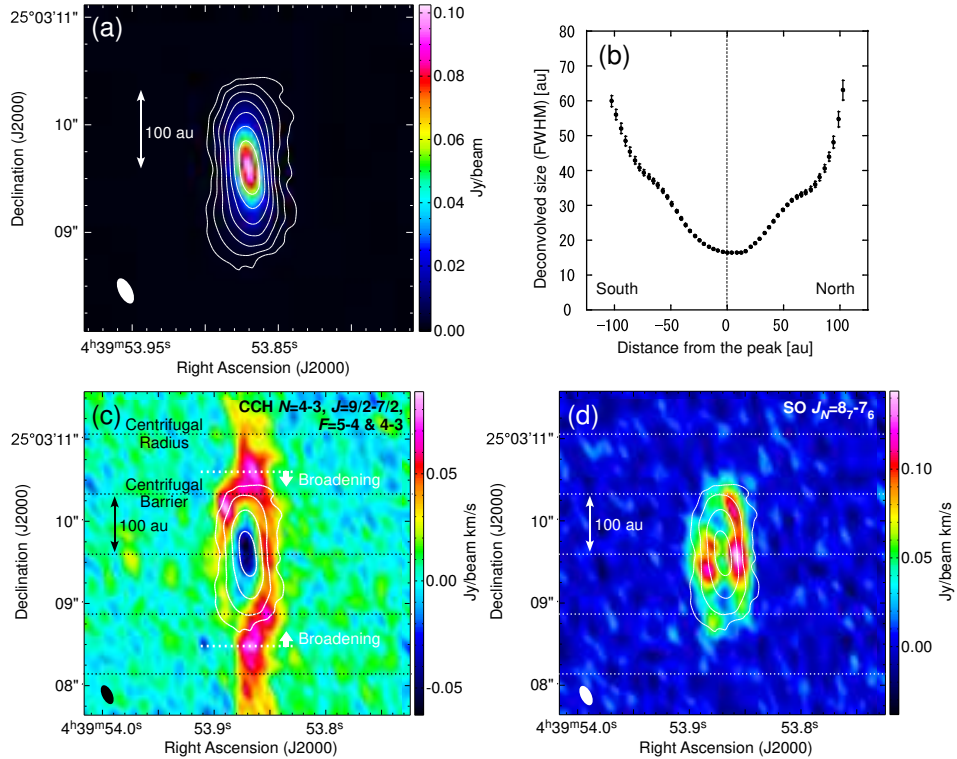}
      \caption{(a) 0.8 mm dust continuum map of L1527.  Contours are 10, 20, 40, 80, 160, 320, and 640 $\sigma$, where 1$\sigma$ is 0.9 mJy beam$^{-1}$  (b) Deconvolved disk thickness as observed in the dust continuum map.  The widths are obtained by the Gaussian fit of the intensity profile along east-west direction.  (c,d) Integrated intensity distributions of CCH ($N=4-3$, $J=9/2-5/2$, $F=5-4$ and $4-3$: colour) and SO ($J_N=8_7-7_6$: colour) superposed on the 0.8 mm dust continuum map (contours: 10, 40, 160, 320, and 640 $\sigma$).  The IRE traced by CCH is broadened inward of the radius of about 150 au.}
         \label{fig:integ}
   \end{figure*}

The 0.8~mm continuum map shows a disk-like structure in an almost edge-on configuration (Figure~\ref{fig:integ}a) \citep{tob13, sak14a, oya15}.  The peak flux is determined to be $96.71\pm0.08$ mJy beam$^{-1}$ from the 2D Gaussian fit, while the integrated flux is $410.6\pm0.4$ mJy.  The integrated flux is 15 \% lower than that observed by ALMA Cycle 0 at the same frequency ($485\pm7$ mJy) \citep{sak14a}. This is probably due to a resolving-out effect and/or a calibration error.  The full-width-half-maximum (FWHM) size of the distribution is $0.62^{\prime\prime} \times 0.21^{\prime\prime}$ with PA of 6.1$^\circ$.  A deconvolved size is $0.57^{\prime\prime} \times 0.15^{\prime\prime}$ (78 au $\times$ 21 au at a distance of 137 pc) with PA of 2.9$^\circ$.  Thus, the continuum emission is resolved.   As a result, the thickness of the disk-like structure is found to vary from the protostar to the northern and southern edges (Figure~\ref{fig:integ}b).  The FWHM thickness of the disk-like structure is measured by the Gaussian fit, as a function of the radial offset from the protostar.  The thickness deconvolved with the synthesized beam is 0.12$^{\prime\prime}$ (16 au) at the continuum peak ($\delta$DEC=0), and increases non-linearly toward both ends ($\sim$100 au). In other words, the disk-like structure is flared within a radius of 100 au.  More interestingly, the thickness versus radius curve has two inflection points at $\pm$70 au from the protostar.  This implies that some structure changes and/or some changes in dust properties (e.g. aggregation of dust grains and/or loss of ice mantle) are occurring there. We leave a detailed understanding on its origin for future studies.  It is interesting to note that the size of the L1527 Keplerian disk was suggested to be 50 au, based on the rotation-curve analysis of the C$^{18}$O emission (Ohashi et al. 2014).
%

\subsection{CCH}
Figure~\ref{fig:integ}c is the moment 0 map of CCH, where the vertical structure of the inner IRE is well resolved.  Along the mid-plane, the CCH emission disappears inward a radius of 100 au.  This radius corresponds to the centrifugal barrier of the IRE according to the ballistic model which successfully reproduces the kinematic behavior of the IRE emission inside a radius of about 1,000 au \citep{sak14a,oya15}.  While the apparent envelope thickness of the IRE is as thin as 40-60 au (FWHM) outward of about 150 au, the envelope thickness is abruptly increased by a factor of 2 inward of it.  Since the centrifugal radius is twice the radius of the centrifugal barrier, the above threshold radius for the broadening is midway between the centrifugal radius (200 au) and the radius of the centrifugal barrier (100 au).  Considering that the CCH emission is absent inside the centrifugal barrier, the broadened region would have a ring-like structure around the protostar.
   \begin{figure*}
   \centering
   \includegraphics[height=13.0 cm]{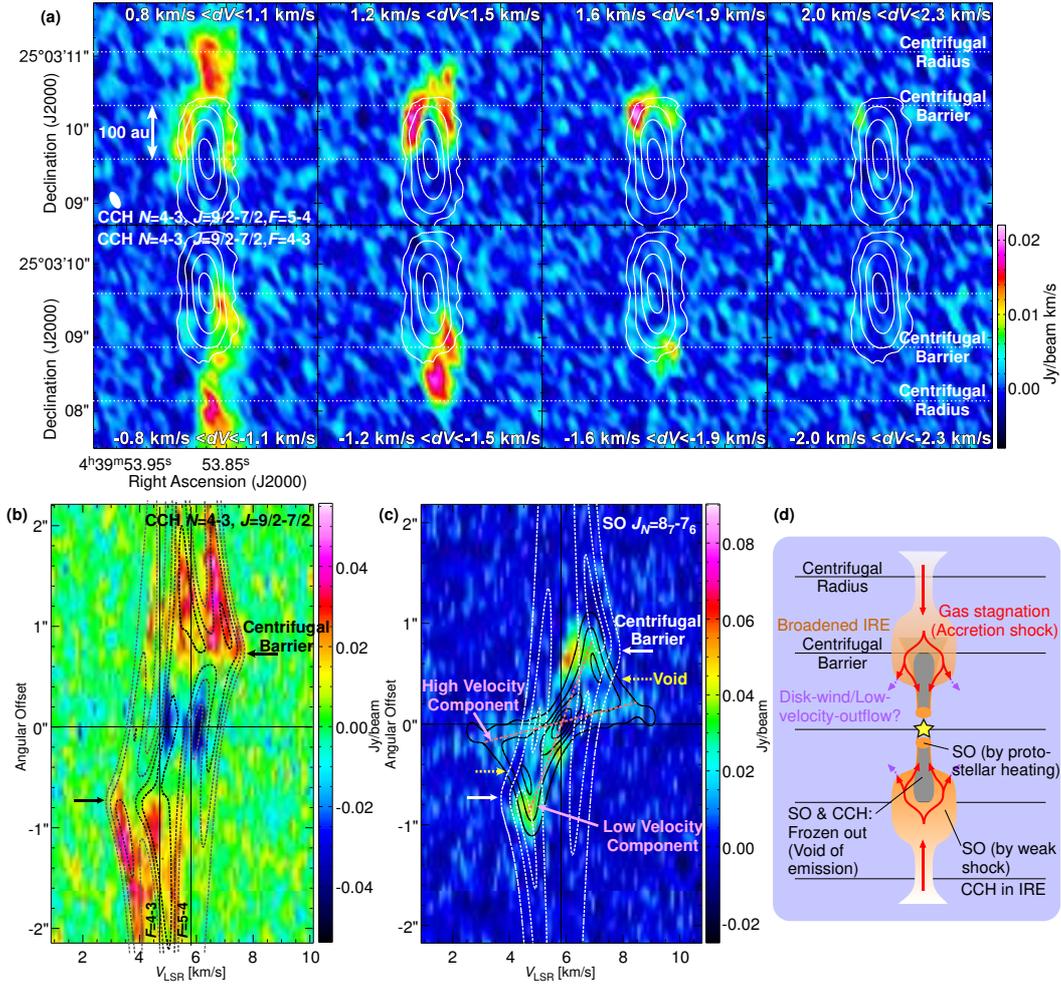}
      \caption{(a) Integrated intensity maps of the CCH ($N=4-3$ $J=9/2-5/2$ $F=5-4$: Upper panels, $N=4-3$ $J=9/2-5/2$ $F=4-3$: Lower panels) line.  Each map is obtained by integrating the velocity range described at bottom in the panel.  (b) Position-velocity (PV) diagram of the CCH lines (colour) superposed to the IRE model (dotted contours).  The model contours are every 20 \% of the peak intensity.  (c) PV diagram of the SO line ($J_N=8_7-7_6$: colour) superposed to the IRE model (dotted white contours) and the Keplerian model (black contours).  Note that the small undulation of the lowest contour of the Keplerian model within 0.2$^{\prime\prime}$ from the centre is an artifact due to a limited mesh size of the model.  In addition to the component associated with the centrifugal barrier, a faint high velocity component can be seen (pink dotted lines).  Yellow dotted arrows indicate the voids of the emission.  (d) Schematic illustration of the physical and chemical structure of the L1527 envelope/disk system.
              }
         \label{fig:vel}
   \end{figure*}
%

Figure~\ref{fig:vel}a shows the velocity channel maps of CCH.  In the $\Delta V$=0.8 - 1.1 km s$^{-1}$ and $\Delta V$=1.2 - 1.5 km s$^{-1}$ panels, the CCH distribution essentially shows a $^{\lq}$Y$^{\rq}$ shaped feature, although significant east-west asymmetry is seen.  The branching point of the $^{\lq}$Y$^{\rq}$ shape corresponds to the centrifugal barrier.  The $\Delta V$=1.2 - 1.5 km s$^{-1}$ component traces the envelope closer to the protostar than the $\Delta V$=0.8 - 1.1 km s$^{-1}$ component.  On the other hand, the $\Delta V$=1.6 - 1.9 km s$^{-1}$ component, which corresponds to the rotation velocity at the centrifugal barrier (1.8 km s$^{-1}$), shows a $^{\lq}$U$^{\rq}$ shaped feature within a radius of 150 au.  No significant emission is seen for $\Delta V$ of 2.0 km s$^{-1}$ or higher, which exceeds the rotation velocity at the centrifugal barrier.  This is consistent with the distribution of the CCH emission, which is absent inside the centrifugal barrier.  In these velocity channel maps as well as the moment 0 map (Figure~\ref{fig:integ}), it looks as if a part of the CCH emission is distributed in the surface area on the disk-like structure traced by the dust continuum emission.  However, this would not be the case, considering the above velocity structure.  This component most likely represents a projection of the ring-like broadened distribution of the infalling-rotating envelope around the centrifugal barrier.  Inside the centrifugal barrier, the density is 10${^8}$ cm$^{-3}$ or higher.  Since CCH is chemically very reactive, it could be lost through gas-phase reactions.  Alternatively, CCH could be depleted onto the dust grains.  Besides, the production of CCH is ineffective in high density gas, because of a low ionization degree (i.e. a low abundance of C$^{+}$).  

Figure~\ref{fig:vel}b shows the position velocity (PV) diagram of CCH along the mid-plane overlaid to the IRE model \citep{sak14b,oya15}.  The IRE model assumes a constant molecular abundance, and does not consider excitation effects.  Nevertheless, the model well reproduces the characteristic kinematic behaviour of the observed PV diagram.  Note that the two hyperfine components are partly blended, which makes the PV diagram complicated.  We also observed the $N$=3-2 transition of CCH and the $J$=5-4 transition of CS with ALMA.  Their behaviour is essentially similar to that of the $N$=4-3 transition of CCH, although the angular resolution is slightly lower.

Here, we note that the distribution of the CCH emission is unaffected by self- and dust-absorption. On the other hand, since a self-absorption feature due to foreground gas is seen at the systemic velocity (Figure~\ref{fig:vel}b), the absorption feature toward the protostar position in the moment 0 map (Figure~\ref{fig:integ}c) is likely due to this foreground gas too.  However, the absence of CCH emission inside the centrifugal barrier is not ascribed to the self-absorption from the foreground gas.  If CCH were present in the mid-plane inside the centrifugal barrier, its emission should be Doppler-shifted due to the rotation motion around the protostar to escape from the self-absorption.  The dust opacity at 345 GHz is evaluated to be less than 0.1 at 100 au, assuming the temperature model for this source reported by \cite{tob13}.  Hence, the absence of the CCH emission is not due to the high opacity of the dust continuum emission, either.  Indeed, the SO line emission distribution is seen even in the closest vicinity ($\sim$10 au) of the protostar, as described later in Section 3.3.

\subsection{SO}
Figure~\ref{fig:integ}d shows the moment 0 map of the SO line.  The SO emission shows complex clumpy structures.  The total flux is 2.0 Jy km s$^{-1}$.  It is 20 \% lower than that in the Cycle 0 observation (2.5 Jy km s$^{-1}$).  The resolving-out effect could thus be seen, but is not serious.  

It is confirmed that the SO emission is absent in the IRE except for the vicinity of the centrifugal barrier.  The SO emission is seen inside the broadened region of the CCH emission.  In order to study the kinematics of the SO containing gas, we prepared the PV diagram along the mid-plane, as shown in Figure~\ref{fig:vel}c.  A weak high-velocity component not detected in the CCH emission is seen near the protostar.  Hence, the PV diagram shows an $^{\lq}$X$^{\rq}$ shaped feature, consisting of the two linear structures (the pink dotted lines in Figure~\ref{fig:vel}c).  The low-velocity component is associated with the ring structure around the centrifugal barrier at a radius of 100 au, which was reported previously \citep{sak14a}.  In Figure~\ref{fig:vel}c, the white contours represent the IRE model.  Although the low-velocity component traces the centrifugal barrier in its position ($\sim$100 au), the rotation velocity is lower than those of the IRE model and CCH.  In contrast, another component appears at higher velocity, which would originate from the inner disk (one of the pink dotted lines in Figure~\ref{fig:vel}c).
  To examine its velocity structure, we calculated the velocity field of the Keplerian motion by using the mass (0.18 M$_{\odot}$) derived from the kinematics of the IRE \citep{sak14a}, as shown by the black contours in Figure~\ref{fig:vel}c.  Here, the outer radius is arbitrarily set to be 160 au.  Although the estimated velocity is still higher than the observed one in some parts, the observed PV diagram seems to be reproduced reasonably well, especially for its high velocity components. 

Apart from the high velocity component detected along the mid-plane, the SO emission looks bright on the surface of the disk-like structure traced by the continuum emission, as shown in Figure~\ref{fig:integ}d.  However, the velocity of this SO emission does not exceed the maximum rotation velocity at the centrifugal barrier, and it does not mainly come from the surface of the Keplerian disk, as in the case of CCH.  It could trace the projection component of the broadened ring-like structure with the clumpy nature around the centrifugal barrier.

As pointed out by Sakai et al. (2014a), it is most likely that SO is liberated into the gas phase from dust grains by a soft accretion shock around the centrifugal barrier.  This picture is justified by the temperature raise ($T\geq$60 K) above the sublimation temperature of SO (50-60 K) \citep[e.g.][]{aot15} at the centrifugal barrier.  In this study, we evaluated the gas kinetic temperature at a higher angular resolution by using the two SO lines, $J_N$=$7_8-6_7$ and $7_6-6_5$.  At this scope, we adopted a non-local thermodynamic equilibrium (non-LTE) method based on the large velocity gradient (LVG) approximation. We used the RADEX code \citep{van07}, where the H$_2$ density and the line width were assumed to be 10$^8$ cm$^{-3}$ and 0.5 km s$^{-1}$, respectively \citep{sak14a,sak14b}.  The temperature is derived for two circular areas with a diameter of 0.4$^{\prime\prime}$ centred at the radius of 100 au and 150 au from the protostar, in the northern part of the disk/envelope system, where the S/N ratio of the spectrum is good enough.  The result is 194 ($^{+146}_{-60}$) K at 100 au and 29($^{+26}_{-11}$) K at 150 au.  As shown in Figure~\ref{fig:integ}d, SO emission is enhanced at the centrifugal barrier, and then decreases again on the mid-plane inside the centrifugal barrier.  This decrease of the SO line intensity corresponds to the positions (30 au $<r<$ 100 au) indicated by the yellow dotted arrows in the PV diagram (Figure~\ref{fig:vel}c).  Unfortunately, the temperature cannot be derived there because of the poor S/N ratio, but an estimation based on the luminosity of the protostar is available \cite{tob13}.  This gives 35 K and 30 K at the radius of 70 au and 100 au, respectively.  The depletion time scale onto the dust grain mantles is very short in dense regions (typically about 1 to 100 years for the density of 10$^{7}$ to 10$^{9}$ cm$^{-3}$) \citep[e.g.][]{aik12}.  Therefore, SO would be re-depleted onto dust grains after passing through the centrifugal barrier. Note that the optical depth of the lines are lower than 0.3, and hence, the optical-depth effect is not needed to be considered here.  Then, the SO emission appears again at the radius of 30 au in the mid-plane.  This would be caused by the thermal evaporation of SO, because the temperature is estimated to be 50$-$70 K there. 

\subsection{Discussion}
The most notable finding in this observation is the broadened inner edge of the IRE located just outside the centrifugal barrier (Figure~\ref{fig:integ}c).  Since the gas cannot penetrate the centrifugal barrier without losing angular momentum, the infalling gas most likely stagnate in front of the centrifugal barrier and causes a weak accretion shock.  Indeed, the temperature raises up to $\sim$190 K (with a large error, see above) in the broadened region, as shown in the analysis of the SO lines.  According to the hydrostatic model for a disk \citep[c.f.][]{har09}, the thickness of the IRE is proportional to $T^{1/2}r^{3/2}$, where $T$ represents the temperature at the radius $r$.  Hence, the temperature rise at a given radius will cause the broadening of the IRE.  Furthermore, a part of the gas would flow out in the vertical directions, which may also contribute to the broadening, as mentioned below.  Above all, we confirm the complex physical structure around the centrifugal barrier, which is accompanied by the drastic change in the gas-phase chemical composition.  Note that in the model of the rotation curve of L1527 by \cite{oha14}, the infalling velocity is artificially reduced inside the radius of $\sim$120 au in order to account for the observed PV diagram of C$^{18}$O.  It corresponds to the stagnant region in front of the centrifugal barrier observed in the present study.
   
In the ballistic model, the velocity at the centrifugal barrier is higher than the Keplerian velocity at the same radius by a factor of $\sqrt[]{\mathstrut 2}$, and hence, some physical mechanisms are expected to fill this gap in the rotation velocity.  In this study, we obtained important hints.  The thickness of the IRE is found to be broadened in front of the centrifugal barrier.  Moreover, the SO molecule, which selectively traces the shocked gas, shows a lower rotation velocity at the centrifugal barrier than the maximum rotation velocity expected for the IRE (Figure~\ref{fig:vel}c).  Most likely, a fraction of the kinetic energy is converted into thermal energy due to the shock, and a fraction of the angular momentum of the SO emitting gas is extracted around the centrifugal barrier.  One possibility is that a fraction of the gas is moving away from the mid-plane in front of the centrifugal barrier, and it might carry away a substantial fraction of angular momentum, as in the so-called $^{\lq}$disk winds$^{\rq}$ or $^{\lq}$low-velocity molecular outflows$^{\rq}$ launched from the centrifugal barrier \citep[c.f.][]{mac11,cod16}.  Further observations examining this hypothesis better are necessary to draw definitive conclusions.

The presented observations demonstrate the importance of the complex vertical structure in the vicinity of the protostar, an aspect that has not been fully studied so far.  The transition from the IRE to the Keplertian disk is much more complex than that previously thought.  High-angular resolution observations will open a new avenue to the comprehension of disk forming regions.

\section*{Acknowledgements}
This paper makes use of the following ALMA data: ADS/JAO.ALMA\#2013.0.00858.S. ALMA is a partnership of ESO (representing its member states), NSF (USA) and NINS (Japan), together with NRC (Canada), NSC and ASIAA (Taiwan), and KASI (Republic of Korea), in cooperation with the Republic of Chile. The Joint ALMA Observatory is operated by ESO, AUI/NRAO and NAOJ.  This study is supported by KAKENHI (25400223, 25108005, and 16H03964).  The authors also acknowledge financial support by JSPS and MAEE under the Japan-France integrated action programme.












\bsp	
\label{lastpage}
\end{document}